\documentclass[a4paper,11pt]{article}
\usepackage{jheppub} % for details on the use of the package, please see the JINST-author-manual
\usepackage{lineno}
\usepackage[dvipsnames]{xcolor}

\usepackage{caption}
\usepackage{subcaption}
\usepackage{mathtools}

\usepackage{slashed}
\usepackage{cleveref}
\usepackage[normalem]{ulem}
\usepackage{multirow}

\usepackage{tablefootnote} 

\usepackage{cancel}

\renewcommand{\d}{\mathrm{d}}
\newcommand{\be}{\begin{equation}}
\newcommand{\ee}{\end{equation}}

\title{\boldmath Neutrino oscillations and scattering theory}

\preprint{P3H-25-027, TTP25-012}

\author[1]{Ilian Dobrev}
\author[1]{Kirill Melnikov}
\author[2]{Thomas~Schwetz}
\affiliation[1]{Institute for Theoretical Particle Physics (TTP), Karlsruhe Institute of Technology (KIT), \\D-76131 Karlsruhe, Germany}
\affiliation[2]{Institute for Astroparticle Physics (IAP), Karlsruhe Institute of Technology (KIT),\\ Hermann-von-Helmholtz-Platz 1, 76344 Eggenstein-Leopoldshafen, Germany}

\emailAdd{ilian.dobrev@student.kit.edu}
\emailAdd{kirill.melnikov@kit.edu}
\emailAdd{schwetz@kit.edu}

{\abstract{We derive the neutrino oscillation probability in vacuum using scattering theory methods  developed earlier  in the context of collider physics~\cite{
  Ginzburg:1991mv,
  Ginzburg:1991mw,
  Kotkin:1992bj}.
It is computed from  Feynman diagrams that
combine neutrino production and detection  processes into a single quantum amplitude.  
Initial-state
particles in the neutrino source and the detector are treated as wave packets.  In contrast to many other approaches, we work with transition probabilities, rather than the  amplitude itself, and  do not specify the form of the wave packets to arrive at  the neutrino oscillation formula. Our approach  offers a simple and transparent framework to discuss decoherence effects in neutrino oscillations, as well as the effects of the finite lifetime of the neutrino source. The latter are particularly relevant for oscillation experiments using neutrinos from pion decays in flight.}


}

\begin{document}
\maketitle
\flushbottom

\section{Introduction}

The phenomenon of neutrino oscillations is, by now, a textbook subject, see e.g.\ 
refs.~\cite{giunti,Suekane:2015yta}. Nevertheless, 
discussions of its proper treatment and interpretation both in quantum mechanics and, especially,  in quantum field theory (QFT) continue, see for instance refs.~\cite{Kayser:1981ye,Giunti:1991ca,Giunti:1993se,Grimus:1996av,Kiers:1997pe,Ioannisian:1998ch,Beuthe:2001rc,Cohen:2008qb,Falkowski:2019kfn,Grimus:2019hlq,Naumov:2020yyv,Karamitros:2022nnh}.  This  reflects an  
obvious fact that neutrino oscillations is a macroscopic phenomenon, whereas the apparatus of
quantum field theory was developed to address physics of microscopic origin, and many familiar concepts  in QFT, such as cross sections and decay rates, 
 are designed to serve this purpose. 
In principle, from the point of view of  quantum field theory,  the description of neutrino 
 oscillations should be straightforward. One considers a single scattering amplitude where 
 a neutrino is first produced in a particular weak decay process, then propagates and eventually collides with 
 a ``detector particle'' producing  a final state which allows its   flavour identification.  
 
However, relating the  scattering amplitude  and the neutrino oscillation probability  must involve a modification of conventional computational methods. Indeed,   this   probability depends on the  \emph{distance} between the source of neutrinos and the detector, and  distances do not appear in  calculations of standard  cross sections in QFT. 
  This conundrum  was resolved thirty years ago, when  it was recognized \cite{Giunti:1993se,Kiers:1995zj,Grimus:1996av,Giunti:1997sk,Giunti:1997wq} that one should employ wave packets for external particles to describe neutrino oscillations in QFT.
Unfortunately, since working with wave packets is 
significantly more involved than with plane waves,  one often  attempts to simplify calculations of the quantum amplitude by  assuming a particular (usually Gaussian) form of  wave packets. Although this approach leads to important insights into the physics of neutrino oscillations,  it  may hide  the generality of the obtained results
\cite{Grimus:1996av,Ioannisian:1998ch,Giunti:1993se,Grimus:2019hlq,Kiers:1997pe,Cohen:2008qb,Naumov:2010um,Naumov:2020yyv,Karamitros:2022nnh}. 

In this respect, it is interesting to note  that there exist  earlier examples  from collider  physics
\cite{Ginzburg:1991mw,Ginzburg:1991mv,Kotkin:1992bj}, where 
 macroscopic (environmental)
conditions  are seamlessly blended into  
 computations of probabilities 
 and (generalized) cross sections.
The  goal  
of refs.~\cite{Ginzburg:1991mw,Ginzburg:1991mv,Kotkin:1992bj} was to develop a mathematical framework to describe radiation of soft photons in collisions 
 of bunches of elementary particles. 
 Since  soft-photon radiation  is  a long-distance phenomenon,  there is 
 a certain wave length where a transition occurs from  radiation in pair-wise collisions of individual constituents of  colliding beams, 
 to   radiation by a    particle from one beam,  scattering on the other beam  as \emph{a whole}. 
 The description of this 
 \emph{beamstrahlung} effect
 in the context of Quantum Electrodynamics (QED) required the development of a particular mathematical apparatus, suitable for separating long-distance and short-distance 
 effects in such processes.  This mathematical apparatus was used later to describe 
 processes at a muon collider whose ``cross sections'' exhibit \emph{linear} sensitivity 
 to the size of colliding beams \cite{Melnikov:1996na} (see also ref. 
 \cite{Dams:2002uy}).  
 
Below we will use these methods from scattering theory to offer a straightforward and transparent approach to deriving  the neutrino oscillation probability. We need neither to specify the form of the wave packets for the initial-state particles nor to introduce the concept of neutrino wave packets, since neutrinos are treated as internal particles in Feynman diagrams.
We will recover an intuitive expression for the overall transition rate in terms of the neutrino production rate, the flavour-transition probability and the detection cross section, convoluted with semi-classical phase-space densities of source and detector particles. The latter allow natural description of  the time structure of source and detector, with phenomenologically interesting applications to  pulsed neutrino beams, or (effectively) stationary detector particles. In addition, we can clarify the   conditions under which  
neutrino oscillations occur either in space or in time. 

We will discuss two further   applications of this computational framework. First, it allows for a straightforward analysis of the conditions which would make the observation of neutrino oscillations impossible. These  so-called decoherence effects 
 represent an important discussion topic in the current literature, see e.g.,  
\cite{Jones:2022hme,Krueger:2023skk,Naumov:2020yyv,Akhmedov:2022bjs,Cohen:2008qb}. 

Second, our framework provides a natural setting to include 
the finite lifetime of the particle whose decay produces the neutrino \cite{Grimus:1999ra,Akhmedov:2010ms,Grimus:2019hlq,Krueger:2023skk,Grimus:2023ktd}. In this respect we identify two effects, namely the smearing of neutrino oscillations due to the finite propagation distance of the decaying particle (in case of a ``small'' width) and the damping of oscillations due to the energy smearing because of the off-shellness of the decaying particle (in case of a ``large'' width). These considerations are relevant for neutrino experiments based on the pion decay in flight, see refs.~\cite{Hernandez:2011rs,Akhmedov:2012uu,Jones:2014sfa} for previous discussions.

The remainder of the paper is organized as follows. In \cref{sec:deriv} we introduce our method and derive the standard formula for the neutrino oscillation probability.  In \cref{sec:decoh} we discuss decoherence effects and the impact of the convolution with the phase-space densities of source and detector particles. In \cref{sec:resonance} we extend the framework to account for the finite lifetime 
and the off-shellness of the decaying source particle. We summarize our findings in \cref{sec:summary}, providing  further discussion of our results and comparing them with the previous literature on the subject. 
Supplementary details about  the resonance calculations are given in \cref{app:resonance}.

%%%%%%%%%%%%%%%%%%%%%%%%%%%%%%%%%%
\section{Derivation of the neutrino oscillation probability}
%%%%%%%%%%%%%%%%%%%%%%%%%%%%%%%%%%
\label{sec:deriv}

For simplicity, we will consider the case of two neutrino flavours $\nu_e$ and $\nu_\mu$.  We will assume that these two neutrino  flavours are linear combinations of two mass eigenstates
$\nu_{1,2}$ with the mixing angle $\theta$
\be
\left (
\begin{array}{c}
  \nu_e \\
  \nu_\mu
\end{array}
\right )
=
\left (
\begin{array}{cc}
\cos \theta  & \sin \theta  \\
- \sin \theta & \cos \theta 
\end{array}
\right )
\;
\left (
\begin{array}{c}
  \nu_1 \\
  \nu_2
\end{array}
\right ). 
\ee
We then express the relevant part of the electroweak 
Lagrangian that  involves interaction of neutrinos
and two  charged leptons ($e,\mu$) as follows 
\be\label{eq:LCC}
\begin{split} 
{\cal L} & = \frac{g}{ \sqrt{2}} W^{-\mu}
\Big [ \bar e \gamma_\mu  P_L  ( \cos \theta \; \nu_1 + \sin \theta \; \nu_2)
%\\& 
+ \bar \mu \gamma_\mu  P_L
( - \sin \theta \; \nu_1 + \cos \theta \; \nu_2) 
\Big ] + {\rm h.c.},
\end{split} 
\ee
where $P_L = (1- \gamma_5)/2$.

\begin{figure}
\centering
\includegraphics[width=0.8\linewidth]{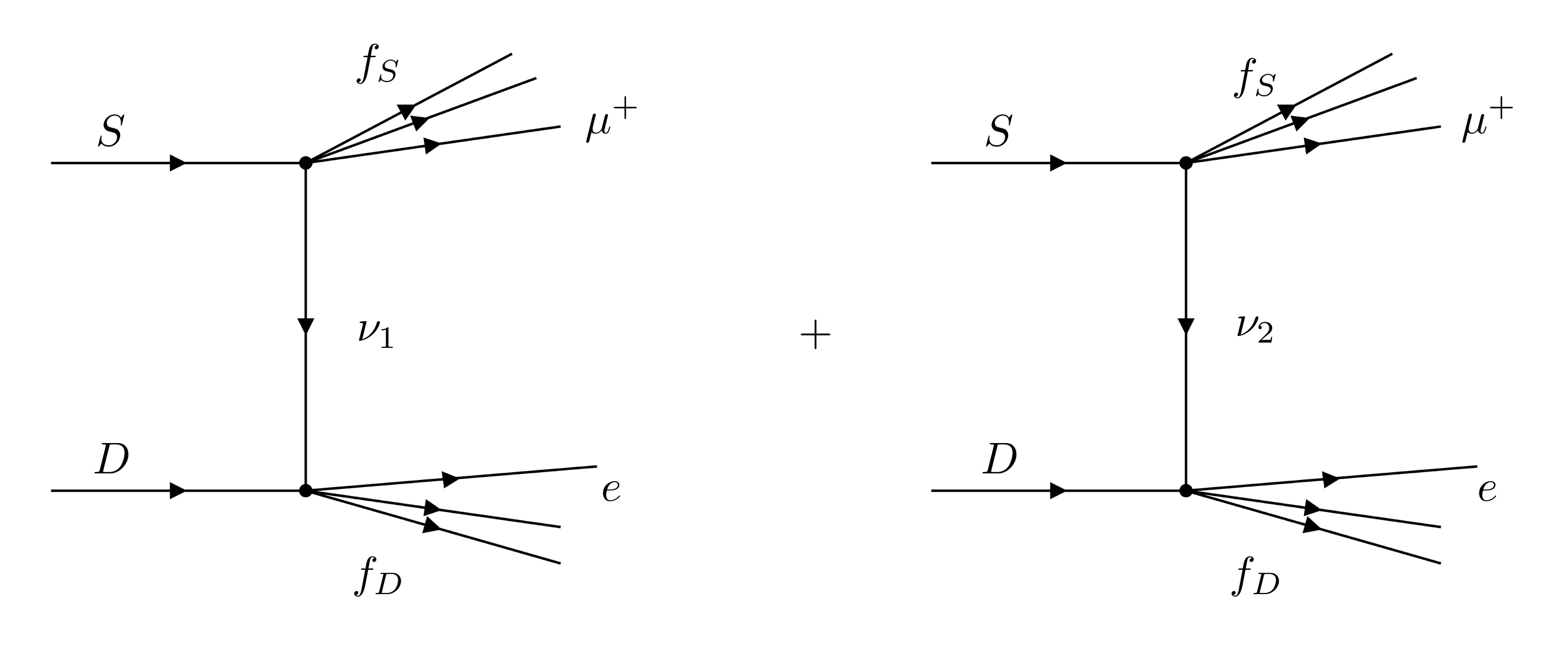}
\caption{\label{fig1}The physical process responsible for neutrino oscillations. Its  
amplitude is the sum of \emph{two} diagrams  with neutrinos $\nu_1$ and $\nu_2$ propagating 
in the $t$-channel.}
\end{figure}

Inspired by the collider analogy, we consider the phenomenon dubbed ``neutrino oscillations'' as a process where a source particle $S$ interacts  with a detector particle $D$  through a  neutrino exchanged in the $t$-channel producing final states $f_{S,D}$
\begin{equation}\label{eq:process}
    S + D \to f_S + f_D \,.
\end{equation}
The process is shown 
in fig.~\ref{fig1}, where 
neutrino mass eigenstates
propagate in the $t$-channel. Both interaction vertices happen via the charged-current weak interaction from \cref{eq:LCC}. The corresponding
4-momenta are denoted by $k_S, k_D$ for $S,D$ and $P_S, P_D$ for $f_{S,D}$.  The final states $f_S$ and $f_D$ both include a charged lepton and we are interested in the transition probability for the full process when flavours of the two charged leptons are different.\footnote{ Modifications to describe a disappearance process with the same flavour are straightforward.} This probability depends on the (macroscopic)
distance between the neutrino production point in the source and the detection vertex, and exhibits an oscillatory pattern \cite{Pontecorvo:1957cp,Bilenky:1978nj}; hence, the name ``neutrino oscillations''. The oscillations emerge from the interference of two diagrams, shown in  fig.~\ref{fig1}. 

It is well-appreciated by now \cite{Grimus:1996av,Beuthe:2001rc,Giunti:1993se,Grimus:2019hlq,Kiers:1997pe,Naumov:2010um,Naumov:2020yyv} that  establishing a connection between the scattering amplitude and the   
neutrino oscillation probability, 
requires  introduction of wave packets for particles 
$S$ and $D$ in the initial state. Quite often wave packets for the final-state particles are also introduced but, as we will see, this is not necessary  and we will not do that 
in what follows, see also \cite{Grimus:1996av, Falkowski:2019kfn, Grimus:2019hlq}.

We describe  initial states at the source and the detector  as superpositions of momentum eigenstates
\be
   |S \rangle  = \int [{\rm d} \vec k_S] \; \phi_S(\vec k_S) \;  | \vec k_S \rangle \,, \qquad
  | D \rangle   = \int 
  [{\rm d} \vec k_D]\; \phi_D(\vec k_D) \;  | \vec k_D \rangle \,,
\ee
where
\be
[{\rm d} \vec k_X] = \frac{ {\rm d}^3 \vec k_X}{(2 \pi)^{3/2} \sqrt{2 E_X} } \,,\;\;\;\;\; X=S,D, 
\ee
and $E_X = \sqrt{\vec k_X^2 + m_X^2}$ is the energy  of the particle 
with the  momentum $\vec k_X$ and the mass $m_X$.
We assume the standard relativistic normalization of 
states $|\vec k_{S,D} \rangle$, i.e., 
\be
\langle \vec k_X' | \vec k_X 
\rangle = 2E_X (2\pi)^3 
\delta^{(3)}(\vec k_X - \vec k_X') \,, 
\ee
and 
\be
\int {\rm d}^3 \vec k_X \;
\phi_{X}(\vec k_X) \;\phi^*_{X}(\vec k) = 1\,,
\qquad X = S,D\,.
\ee
It follows that 
\be
\langle S | S \rangle = \langle D | D \rangle =1.
\ee

The $S$-matrix element for the process \cref{eq:process} is given by
\begin{equation}
  S_{if} = \int 
 [{\rm d} \vec k_S]
 [{\rm d} \vec k_D]
 % \frac{\d^3k_S\d^3k_D}{(2\pi)^3\sqrt{2E_S 2E_D}} 
 \phi_S(\vec k_S)\phi_D(\vec k_D)
  (2\pi)^4\delta^{(4)}(k_S + k_D - P_S - P_D) \mathcal{M}_{if} \,.
\end{equation}
Then the differential transition probability $i\to f$ is obtained from the following equation:  
\be
\d W_{if} = |S_{if}|^2 \;\d\nu_f \,, 
\ee
where ${\rm d} \nu_f$ is the density of final states.
%and we  assumed that initial and final states are normalized to one. 
It reads
\begin{equation}
  \d\nu_f = \d\Phi_S \;\d\Phi_D
  \,,\qquad \d\Phi_S = \prod_{x\in S} \frac{\d^3p_x}{(2\pi)^32E_x} 
  \,,\qquad \d\Phi_D = \prod_{x\in D} \frac{\d^3p_x}{(2\pi)^32E_x} \,,
\end{equation}
where $x$ runs over all 
final-state particles, either in the source or in the detector. We find 
\be
\begin{split}
  \d W_{if} =& \int 
  [{\rm d} \vec k_S]
  [{\rm d} \vec k_S']
  [{\rm d} \vec k_D]
  [{\rm d} \vec k_D']
  \;  \d\Phi_S \d\Phi_D \,
  \phi_S(\vec k_S)\phi_D(\vec k_D)\phi_S^*(\vec k'_S)\phi_D^*(\vec k'_D)  \\
  &
  \times (2\pi)^4\delta^{(4)}(k_S + k_D - P_S - P_D) ( 2\pi)^4 \delta^{(4)}(k'_S + k'_D - P_S - P_D)
  \mathcal{M}_{if} \mathcal{M}'^*_{if} \,, \label{eq:W1}
\end{split}
\ee
where $\mathcal{M}'_{if}$ indicates that this matrix element depends on the primed momenta.

To simplify this expression, we follow the approach described in 
refs.~\cite{Ginzburg:1991mv,Ginzburg:1991mw,Kotkin:1992bj}, but also introduce an important  modification 
that is needed to describe the set-up of neutrino oscillation experiments.
We start with an observation that for the description of  a sequential process,  where  a neutrino is produced at the source, 
and then  absorbed at the detector,  it is useful to introduce  the neutrino momenta
\be
q = k_S - P_S = P_D - k_D,
\;\;\; q' = k_S'-P_S 
= P_D - k_D'.
\ee
Then, we insert 
\be
1 = \int {\rm d}^4 q 
\delta^{(4)}(k_S - P_S -q) 
 \int {\rm d}^4 q'  
\delta^{(4)}(k_S' - P_S -q'),
\ee
into eq.~(\ref{eq:W1}), 
rewrite delta-functions there separating them into ``production'' and ``detection'' ones,  change variables to 
\begin{align}\label{eq:momenta}
  Q = \frac{q+q'}{2} \,,\quad \kappa = q-q' \,,\qquad
  l_X = \frac{k_X+k'_X}{2} \,,\quad \kappa_X = k_X - k'_X \,,\quad (X=S,D), 
\end{align}
and obtain 
\be
\begin{split}
  & \d W_{if}  = \int 
  [{\rm d} \vec k_S]
  [{\rm d} \vec k_S']
  [{\rm d} \vec k_D]
  [{\rm d} \vec k_D']
  \;  \d\Phi_S \d\Phi_D \,
  \phi_S(\vec k_S)\phi_D(\vec k_D)\phi_S^*(\vec k'_S)\phi_D^*(\vec k'_D)   \mathcal{M}_{if} \mathcal{M}'^*_{if} \\
  &
  \times {\rm d}^4 \kappa  
{\rm d}^4 Q 
\delta^{(4)}(\kappa_S - \kappa) \delta^{(4)}(\kappa_D + \kappa)  
 (2 \pi)^4 \delta^{(4)}(l_S - P_S  - Q)
 (2 \pi)^4 
 \delta^{(4)}(l_D - P_D  + Q) 
 \,, \label{eq:W2}
\end{split}
\ee
where we have also  used $\d^4q\d^4q' = \d^4Q\d^4\kappa$. Physically, $Q,l_S,l_D$ correspond to mean four-momenta of the internal neutrino, the source and the detector particles, respectively, whereas $\kappa,\kappa_{S,D}$ describe the momentum differences in the interfering states. 
%As we will see below, $\kappa,\kappa_{S,D}$ are constrained by the localization of source and detector particles. 
We emphasize again that final-state momenta of all particles are kept fixed. 

We then write the energy part of the 
$\delta$-function involving $\kappa$'s in \cref{eq:W2} as integrals over ``source'' and ``detector'' times 
\begin{align} \label{eq:delta-kappa2}
 (2 \pi)^2  \delta(\kappa^0_S - \kappa^0) \delta(\kappa^0_D + \kappa^0) =
  \int \d t_S \, e^{-i(\kappa^0_S-\kappa^0)t_S}
  \int \d t_D \, e^{-i(\kappa^0_D+\kappa^0)t_D} \,.
\end{align}
As recognized in 
refs.~\cite{Ginzburg:1991mv,Ginzburg:1991mw,Kotkin:1992bj}, a step  leading to   key 
simplifications in  eq.~(\ref{eq:W2}),  
is  the statistical averaging  over the various quantum states described by the wave functions 
$\phi_S$ and $\phi_D$. This averaging produces time-dependent momentum-space
density matrices 
\begin{align}\label{eq:density-matrix}
  \langle\phi_X(\vec k_X)\phi_X^*(\vec k'_X) e^{-i\kappa^0_Xt_X}\rangle  =
  \rho_X(\vec k_X, \vec k'_X, t_X), 
\end{align}
for $X=S,D$. Another crucial observation of 
refs.~\cite{Ginzburg:1991mv,Ginzburg:1991mw,Kotkin:1992bj}  is the utility of the so-called Wigner function \cite{Case2008}, 
related to the 
density matrix through a particular Fourier transform  
\begin{align}\label{eq:density}
  \rho_X(\vec k_X, \vec k'_X, t_X) = \int \d^3 \vec r_X \, n_X(\vec r_X, \vec l_X, t_X) e^{-i\vec\kappa_X\vec r_X}, \, \qquad X=S,D\,.
\end{align}
Indeed, the classical limit of a Wigner function is the  
phase-space distribution of particles described by the relevant density matrix;  in our case, these distributions are
exactly what is needed to describe  the source and the detector of neutrinos.

We note that 
eqs.~(\ref{eq:delta-kappa2}, \ref{eq:density-matrix}, \ref{eq:density})
%\cref{eq:delta-%kappa2,eq:density-%matrix,eq:density} 
show  important difference 
between  discussion in refs.~\cite{Ginzburg:1991mv,Ginzburg:1991mw,Kotkin:1992bj}, which focused on collider processes, and the neutrino oscillations case which we want to describe here.  In refs.~\cite{Ginzburg:1991mv,Ginzburg:1991mw,Kotkin:1992bj}, 
$\kappa^0$ 
was integrated out, and a single time-like variable was introduced $(2\pi)\delta(\kappa^0_S - \kappa^0_D) = \int \d t \, e^{-i(\kappa^0_S-\kappa^0_D)t}$. This implies that  phase-space densities at both vertices are evaluated at the same value of $t$. In contrast, \cref{eq:delta-kappa2} introduces two separate time variables for the source and detector, such that source and detector densities are evaluated at different times. This is important  for properly describing  retardation effects that are essential because  sources and detectors in neutrino oscillation experiments can be separated by very large distances. Hence,   as we will find below, relevant values of $t_S$ and $t_D$ will be determined by specific properties of the source and detector, as well as time required for a neutrino to propagate between them.

Physically, replacing the time components of the delta-functions with the time-dependent densities $n_S(t_S)$ and $n_D(t_D)$ corresponds to introducing energy non-conservation: energy will be conserved in the source/detector processes only within accuracy $\delta E_X \sim 1/\tau_X$, where $\tau_X$ is a characteristic time scale over which source or detector densities $n_{S,D}$ change. Different from collider experiment, in neutrino oscillation experiments the time structure of the source and the detector  are  unrelated and can be very different, making introduction of two, rather than one, times an important  physically-motivated modification. 

Returning to the calculation of 
the transition probability $\d W_{if}$ 
in eq.~(\ref{eq:W1}), we change the integration variables $\d^3 \vec k_X \d^3 \vec k'_X = \d^3 \vec l_X\d^3 \vec \kappa_X$, $X = S,D$, and integrate over  $\d^3 \vec \kappa_S \d^3 \vec \kappa_D$ finding  $\vec \kappa = \vec \kappa_S= -\vec \kappa_D$.
After collecting  remaining terms, \cref{eq:W1} becomes
\begin{equation}
\begin{split}
  \d W_{if} =& \int 
  \prod \limits_{x = {S,D}}^{}
\frac{\d^3 \vec l_X} {\sqrt{2E_X 2E'_X}}\; 
\d t_X   \d^3 \vec r_X\d^3  n_X(\vec r_X, \vec l_X, t_X)\, 
\d^4 Q \d^4 \kappa 
\;   e^{-i\kappa_\mu x^\mu}  
  \\
  & \times 
\delta^{(4)}(l_S - P_S  - Q)\delta^{(4)}(l_D - P_D  + Q)
\mathcal{M}_{if} \mathcal{M}'^*_{if} \,
\d\Phi_S\d\Phi_D 
, \label{eq:W3}  
\end{split}
\end{equation}
where $x^\mu = (t_D - t_S, \vec r_D - \vec r_S)$.
We continue with the discussion of the matrix element and note that for processes that lead to neutrino oscillations, the neutrino in the $t$-channel should be close to its mass shell.  It follows that the amplitude can be written in a factorised form  
\begin{align}
  \mathcal{M}_{if} \approx  -\sin\theta\cos\theta \, M_S\sum_{a=1,2} \frac{(-1)^a}{q^2-m_a^2+i\epsilon} M_D\,.
\end{align}
where, $\theta$ is the mixing angle, and $M_S,M_D$ are the matrix elements for the production and detection processes of the \emph{massless on-shell} neutrinos of the respective flavor, and the sum originates from  the two diagrams with the two neutrino mass eigenstates  in the $t$-channel. Neutrino masses are referred to as  $m_{1,2}$.  Another natural  assumption is that the dependence of $M_{S,D}$ on the neutrino momenta is smooth enough, to neglect  differences between $q$ and $q'$ there and use  $q\approx q' \approx Q$. Then we obtain
\begin{align}
  \mathcal{M}_{if} \mathcal{M}'^*_{if} = 
  \sin^2\theta\cos^2\theta \, |M_S|^2 |M_D|^2\sum_{a,b=1,2} \frac{(-1)^a(-1)^b}
      {(q^2-m_a^2+i\epsilon)(q'^2-m_b^2-i\epsilon)} \,.
\end{align}
We use \cref{eq:momenta},
to re-write  neutrino momenta squared $q^2$ and $q'^2$ in terms of $Q$ and 
$\kappa$  finding  
\be
\begin{split}\label{eq:q-sq}
  q^2 &= (Q + \kappa/2)^2 \approx Q^2  + Q \kappa +{\cal O}(\kappa^2) \,, 
  \\ 
  q'^2 & = (Q - \kappa/2)^2 \approx Q^2 - Q \kappa + {\cal O}(\kappa^2) \,.
\end{split}
\ee
As we will see later, the magnitude of the vector $\kappa$ is determined by the macroscopic distance between source and detector. For this reason $\kappa/Q$ is tiny, and   
we  can neglect terms of order $\kappa^2$ in what follows. 
The transition probability becomes
\be
\begin{split}
  \d W_{if} &= \sin^2\theta\cos^2\theta  \int \frac{ \d^4Q}{(2 \pi)^4 } \\
  &\times \d\Phi_S \int \frac{\d^3 \vec l_S}{2E_S} \int \d t_S \d^3\vec r_S \, n_S(\vec r_S, \vec l_S, t_S) \, (2 \pi)^4 \delta^{(4)}(l_S - P_S  - Q) \, |M_S|^2 \\
  &\times \d\Phi_D \int \frac{\d^3\vec l_D}{2E_D} \int \d t_D \d^3 \vec r_D \,n_D(\vec r_D, \vec l_D, t_D) \, (2  \pi)^4 \delta^{(4)}(l_D - P_D  + Q) \, |M_D|^2 \\
&\times  \int 
\frac{\d^4\kappa}{(2 \pi)^4} \,  e^{- i \kappa x}  
\sum_{a,b=1,2} \frac{(-1)^a(-1)^b}{(Q^2 + Q\kappa -m_a^2+i\epsilon)(Q^2-Q\kappa-m_b^2-i\epsilon)} \,,
  \label{eq:W4}  
  \end{split}
\ee
where we replaced 
$E_{S,D}$ and $E_{S,D}'$ 
with $E_{S,D}(\vec l_{S,D})$.

The main contribution to the above integral comes from the integration region 
where $Q^2 \sim m_{1,2}^2$.
For such $Q^2$, we can 
re-write the integration 
$\d^4Q$ in such a way that the integration over the neutrino ``mass'' $Q^2$ and its phase-space element ${\rm d}^3 \vec Q/(2 Q_0 (2 \pi)^3) $ with 
$Q_0 = \sqrt{\vec Q^2 + Q^2}$ appear, by changing integration variables $\d Q^0\to \d Q^2$.
Furthermore, making the natural assumption that neutrinos produced in the source are ultra-relativistic, we can drop 
$Q^2$ in the expression for 
$Q_0$.
Then,  \cref{eq:W4} 
simplifies as follows
\be
\begin{split}
  \d W_{if} &= \sin^2\theta\cos^2\theta  \int \frac{\d Q^2}{2\pi} 
  \int \frac{\d^3 \vec l_S}{2E_S} \int \d t_S \; \d^3 \vec r_S \, n_S(\vec r_S, \vec l_S, t_S) \,
  \\
  &\times 
   (2\pi)^4\delta^{(4)}(l_S - P_S  - Q) \, |M_S|^2 \d\Phi_{S} 
  \frac{\d^3\vec Q}{(2\pi)^3 2|\vec Q|}\\
  &\times  \int \frac{\d^3 \vec l_D}{2E_D} \int \d t_D \; \d^3 \vec r_D \,n_D(\vec r_D, \vec  l_D, t_D) \, (2\pi)^4\delta^{(4)}(l_D - P_D  + Q) \, |M_D|^2 \d\Phi_D \\
&\times  \int \frac{\d^4\kappa}{(2\pi)^4} \,  e^{- i \kappa x}  
\sum_{a,b=1,2} \frac{(-1)^a(-1)^b}{(Q^2 + Q\kappa -m_a^2+i\epsilon)(Q^2-Q\kappa-m_b^2-i\epsilon)} \,.
  \label{eq:W5}    
\end{split}
\ee
As we already mentioned, 
neutrinos are nearly on-shell,  which implies that in $Q \kappa$ terms,
we replace $Q_0$ with $|\vec Q|$. This is accurate 
%$Q^0 \approx |\vec Q|$, and we have $Q\kappa \approx |\vec Q|\kappa^0 - \vec Q\vec \kappa$, 
up to  $\mathcal{O}(\kappa m_\nu)$ terms, which we neglect.
Then the integral over $Q^2$ is only important for  the last line in  \cref{eq:W5}. 
Hence, we define a function
\be
\label{eq2.26}
I(x_\mu,Q^\mu) = 
\int \frac{{\rm d} Q^2}{2 \pi} \;\frac{\d^4\kappa}{(2\pi)^4} \,  e^{- i \kappa x}  
\sum_{a,b=1,2} \frac{(-1)^a(-1)^b}{(Q^2 + Q\kappa -m_a^2+i\epsilon)(Q^2-Q\kappa-m_b^2-i\epsilon)} \,,
\ee
which, as we will see shortly, provides the neutrino oscillation probability.   To compute this function, we integrate over $Q^2$ using the residue theorem
and find
\be
    I \equiv \frac{i}{|\vec Q|}
  \int \frac{\d^4\kappa}{(2\pi)^4} \,  e^{- i \kappa x}  
\left [ \frac{1}{n_\nu \kappa + i\epsilon}
- \frac{1}{2} 
\left ( \frac{1}{ n_\nu \kappa + 2L^{-1}_{\rm osc}
+ i\epsilon}
+ \frac{1}{ n_\nu \kappa - 2 L_{\rm osc}^{-1}+ i\epsilon} 
\right )  
\right ].
\label{eq.2.27}
\ee
where the neutrino oscillation length is defined as
\be
L_{\rm osc} = \frac{4|\vec Q|}{\Delta m^2},  
\ee
$\Delta m^2  \equiv m^2_2-m^2_1$, and $n_\nu = (1, \vec e_\nu)$ is the four-vector that describes the world-line of a propagating ultra-relativistic neutrino. 

To complete the integration in \cref{eq.2.27}, we  use the identity
\begin{equation}
    \frac{i}{A+i\epsilon } \equiv \int \limits_0^\infty \d\tau \, e^{iA\tau},
\end{equation}
integrate over $\kappa$, and find 
\begin{align}
%    &= \int \limits_0^\infty \d\alpha \int \frac{\d^4\kappa}{(2\pi)^4} 
%     e^{-i\kappa(x - \alpha Q)}
%\left[1 - \frac{1}{2}\left(e^{i2 \alpha/L_{\rm osc}} 
%+
%e^{-i2 \alpha/L_{\rm %osc}} 
%+e^{-i\alpha\Delta m^2/2} 
%\right) \right] 
I = \frac{2}{|\vec Q|} \int \limits_0^\infty \d\tau \, \delta^{(4)}(x - \tau n_\nu) \sin^2 \left( \frac{\tau}{L_{\rm osc}}
\right) \,.
\end{align}
%\left[1 - \cos\left(
%\frac{2 \tau}{L_{\rm osc}}
%\frac{\alpha\Delta m^2}%{2} 
%
To proceed further, we use
the definition of the four-vector $x^\mu = (t_D-t_S, \vec r_D - \vec r_S) $, and 
explicitly write the delta-function by separating 
components of $\vec x$ into parallel and orthogonal ones relative to  the neutrino momentum. We obtain
\be
    \delta^{(4)}(x - \tau n_\nu) = 
    \delta(t_D - t_S - \tau) \,
    \delta(r_{||} - \tau) \,
    \delta^{(2)}( \vec r_\perp
) \,,
\ee
where $\vec r = \vec r_D - \vec r_S$, $r_{||} = 
\vec r \cdot \vec e_\nu$ 
and $\vec r_\perp \cdot \vec e_\nu = 0$.
Integrating over $\tau$,
we find 
\begin{equation}\label{eq:I}
    I = \frac{2}{|\vec Q|} 
    \delta(t_D - t_S - r_\parallel) \,
    \delta^{(2)}(\vec r_{\perp}) \, \Theta(r_{||})
   \sin^2 \left ( \frac{r_{||}}{L_{\rm osc}} \right ) \,.
\end{equation}
Using this result in \cref{eq:W5}, integrating over detection time $t_D$, and replacing the neutrino momentum $Q$ with $p_\nu = E_\nu(1, \vec e_\nu)$ for the sake of clarity, we find 
\be
\begin{split}
  \d W_{if} &= 
  \int \d t_S \; \d^3 \vec l_S    \d^3 \vec r_S 
\, n_S(\vec r_S, \vec l_S, t_S) 
  \d^3 \vec l_D \d^3 \vec r_D \,n_D(\vec r_D, \vec l_D, t_S + ( \vec  r_D - \vec r_S) \cdot \vec e_\nu ) 
  \\
  & \times \frac{1}{2 E_S} (2\pi)^4\delta^{(4)}(l_S - P_S  - p_\nu) \, |M_S|^2 
  \; \d\Phi_{S} 
  \frac{\d^3\vec p_\nu}{(2\pi)^3 2 E_\nu}
  \\
  &\times  \,
  \frac{1}{2E_\nu}  
   \frac{1}{2E_D} \, (2\pi)^4\delta^{(4)}(l_D - P_D  + p_\nu) \, |M_D|^2 \d\Phi_D\\
&\times      \delta^{(2)}( \vec r_{D, \perp} - \vec r_{S,\perp} ) \, 
\Theta[( \vec  r_D - \vec r_S) \cdot \vec e_\nu]
\,
\sin^22\theta \; 
\sin^2\left(\frac{ ( \vec  r_D - \vec r_S) \cdot \vec e_\nu }{L_{\rm osc}}\right) \,.
\label{eq:W6}    
\end{split}
\ee

We can easily identify  the physical meaning of several contributions to the above equation.
The first one is the 
 standard neutrino oscillation probability
\begin{align}
 P_{\rm osc}(\vec r_D, \vec r_S, \vec e_\nu) = 
  \sin^22\theta \sin^2 \left ( \frac{ (\vec r_D - \vec r_S) \cdot \vec e_\nu }{L_{\rm osc}}
  \right ) \,.
\end{align}
We note that, when this expression is used in eq.~(\ref{eq:W6}) together with the transverse delta-function, 
we can replace 
$(\vec r_D - \vec r_S) \cdot \vec e_\nu = |\vec r_D - \vec r_S|$.

The second 
identifiable contribution in \cref{eq:W6} is  
the differential decay rate of the source particle in the laboratory frame 
  \begin{align}
    \d\Gamma_S = \frac{1}{2E_S}  (2\pi)^4\delta^{(4)}(l_S - P_S  - p_\nu) \, |M_S|^2 \, \d\Phi_{S\nu}. \label{eq:Gamma}
  \end{align}
We note that  
\begin{equation}
    \d\Phi_{S\nu} = \frac{\d^3p_\nu}{(2\pi)^32E_\nu}
 \, \d\Phi_S \,,
\end{equation}
is the phase space element  for the final state of the process $S\to f_S + \nu$.

Third, the quantity in the third line of 
  \cref{eq:W6} is related to the differential cross section of the process $D+\nu \to f_D$ 
  \begin{align}   \label{eq:sigma}
    (1 - \vec e_\nu \cdot \vec \beta_D ) \; \d\sigma_{D \nu} = \frac{1}{2E_D2E_\nu } \d\Phi_D (2\pi)^4\delta^{(4)}(l_D - P_D  + p_\nu) \, |M_D|^2 \,,
   \end{align}
where $\vec \beta_D$ is the velocity of the detector particle $D$. 

Finally,  it is possible to  write the transversal delta-function as 
\begin{align}\label{eq:delta-transv}
  \delta^{(2)}(\vec r_\perp) =
  \frac{1}{r^2} \,
  \left ( \delta^{(2)}( \vec e_r -  \vec e_\nu ) 
  + 
  \delta^{(2)}( \vec e_r + \vec e_\nu ) 
  \right ), 
\end{align}
where $\vec e_r = \vec r/r$ and the delta-functions on  the right-hand side are defined as follows  
\be\label{eq:delta-angles}
\int {\rm d} \Omega_{\vec r } \; 
\delta^{(2)}(\vec e_r - \vec e_\nu)
f(\vec r) = 
f(r \vec e_\nu).
\ee
\Cref{eq:delta-transv} leads  to the  inverse distance-squared dependence of the neutrino flux at the detector.    
Hence, the transition probability in \cref{eq:W6} can be written as follows
\be
\begin{split}
  \d W_{if} &= \int \d^3 \vec p_\nu \,
  \int \d t_S \d^3\vec l_S \d^3 \vec r_S \,  n_S(\vec r_S, \vec l_S, t_S) \, \frac{\d\Gamma_S}{\d^3 \vec p_\nu} 
  \\
& \times
\int  \d r \, \d^3 \vec l_D \, n_D(\vec r_S + r \vec e_\nu , \vec l_D, t_S + r) \, \d\sigma_{D \nu} \; P_{\rm osc}(r) \,, \label{eq:W7}
\end{split}
\ee
where we have assumed that the detector particle is non-relativistic and replaced  the factor $1-\vec e_\nu\cdot \vec \beta _D$ in \cref{eq:sigma} with one. Furthermore, we have changed the integration variables from $\d^3 \vec r_D \to \d^3\vec r$, switched to polar coordinates, and used \cref{eq:delta-angles} for the angular integration. The factor $1/r^2$ from \cref{eq:delta-transv} cancels with the $r^2$  from the volume element in   spherical coordinates.

%%%%%%%%%%%%%%%%%%%%%%%%%%%%%%%%%%%%%%%%%%%%%%%%%%%%%%%%%%%%%%%%%%%%%
\section{Impact of the source and detector densities and the decoherence effects}
%%%%%%%%%%%%%%%%%%%%%%%%%%%%%%%%%%%%%%%%%%%%%%%%%%%%%%%%%%%%%%%%%%%%
\label{sec:decoh}

\Cref{eq:W7} offers a very suggestive expression, where the product of the  decay rate for the source particle multiplied with  the oscillation probability and  the detection cross section is convoluted with the phase-space densities of source and detector particles. This convolution 
impacts the possibility to  observe  neutrino oscillations as we will now explain. In addition to these classical averaging effects, the derivation in the previous section allows to identify  assumptions leading to the standard oscillation probability, and quantify  conditions, under which these assumptions hold. Once they are violated, decoherence effects will modify the oscillation pattern. We will comment on this at the end of this section, and show that  these effects are parametrically equivalent to the classical averaging suggested by \cref{eq:W7}.

In order to study this effect, we will  consider a toy model, assuming that the source and detector phase-space densities have a Gaussian shape in space, momentum and time 
\begin{align}
  n_X(\vec r_X, \vec l_X, t_X) \propto \exp\left[
    -\frac{1}{2}\left(\frac{\vec r_X - \vec L_X}{\delta_X}\right)^2 
    -\frac{1}{2}\left(\frac{\vec l_X - \vec {\sf l}_X}{\sigma_X} \right)^2
    -\frac{1}{2}\left(\frac{t_X - T_X}{\tau_X}\right)^2
    \right] \,,
    \label{eq3.1}
\end{align}
for $X = S,D$, with the mean locations $\vec L_{S,D}$, mean momenta $\vec {\sf l}_{S,D}$, and mean times $T_{S,D}$. For simplicity we assume the distributions to be isotropic. Quantum mechanics requires
\begin{align}\label{eq:QM}
\delta_X \sigma_X > \frac{1}{2} \,.  
\end{align}
The time spreads $\tau_{S,D}$ represent  typical time scales, over which the source and detector change. As mentioned above, this implies that energy is conserved in the corresponding processes only approximately, up to $\delta E_X \sim 1/\tau_X$ ($X=S,D$).
In the next section we will  consider a specific realisation of such a situation -- the time dependence induced by the decay of the source particle. 

Let us first consider the consequences of the spatial and time localizations. As the decay rate and detection cross section do not depend on space and time, only the delta-functions and the oscillation probability are folded with the corresponding Gaussians in eq.~(\ref{eq3.1}). Substituting these densities into \cref{eq:W7}, we find that the following integral is needed 
\begin{equation}
\int \limits_{0}^{\infty} \d r\int 
\d t_S \; \d^3 \vec r_S
e^{-\frac{1}{2}\left(\frac{\vec r_S - \vec L_S}{\delta_S}\right)^2} 
e^{-\frac{1}{2}\left(\frac{t_S  - T_S}{\tau_S} \right )^2}
e^{-\frac{1}{2}\left(\frac{\vec r_S + r \vec e_\nu - \vec L_D }
{\delta_D}\right)^2} 
e^{-\frac{1}{2}\left(\frac{t_S +r - T_S}{\tau_D} \right )^2}
P_{\rm osc}(r).
\end{equation}
To compute it, we express the sine function that appears in the neutrino oscillation probability as a linear combination of complex-valued exponential functions, 
extend the integral over $r$ to $-\infty$, which allows us to get leading contributions without having to deal with the error functions,  
and find the following expression for the ``smeared'' oscillation probability
\be
\begin{split}
P_{\rm osc} &= \frac{1}{2} \sin^22\theta 
 \exp\left[-\frac{(\vec L_{D\perp} - \vec L_{S\perp})^2}{2\delta^2} \right]
  \exp\left[-\frac{1}{2} \frac{(L-T)^2}{\tau^2+\delta^2} \right]
  \\
  &\times
  \left[1 - 
  \exp\left(-\frac{1}{2} \frac{\tau^2\delta^2}{\tau^2+\delta^2} \Delta^2\right)
  \cos\left(\Delta \frac{\tau^2 L + \delta^2 T}{\tau^2+\delta^2} \right)
  \right]\,, \label{eq:Pdecoh-space-time}
\end{split}
\ee
where 
\begin{equation}
\Delta \equiv \frac{\Delta m^2}{2E_\nu} 
= \frac{2}{L_{\rm osc}}\,,    
\end{equation}
and 
\be
\begin{split}\label{eq:spreads-space-time}
  &\tau^2 \equiv \tau_D^2 + \tau_S^2 \,,\quad \delta^2 \equiv \delta_D^2 + \delta_S^2 \,,\quad L \equiv (\vec L_{D} - \vec L_{S}) \cdot \vec e_\nu, \,\quad T \equiv T_D - T_S \,.
\end{split}
\ee
The exponential involving $\vec L_{S,D\perp}$  constrains the direction of $\vec p_\nu$ within an opening angle of $\lesssim\delta/L$,
whereas the second exponential in the first line constrains the mean distances $T\approx L$ within $\sqrt{\tau^2+\delta^2}$. The exponential in the second line gives the decoherence suppression of oscillations. It is dominated by the smaller spread  of $\tau$ and $\delta$, i.e., oscillations are damped if $ \rm{min}(\tau, \delta) \Delta \gtrsim 1$. 

The cosine term in eq.~(\ref{eq:Pdecoh-space-time})  describes the modified oscillation phase.  If $\tau \gg \delta$ ($\tau \ll \delta$) oscillations happen in space (time). Hence, for $\tau \lesssim \delta$ we would predict deviations from the canonical oscillations in space.

The limit $\tau \to \infty$, that can be realized by taking $\tau_D \to \infty$, corresponds to the case of a \textit{stationary detector}.\footnote{We note that for  the source we cannot take the limit $\tau_S\to\infty$, as we need the source particle to decay in order to get a neutrino.} Apart from the constraint in 
the transverse direction, we obtain in this case
\begin{align}\label{eq:decoh-stationary}
P_{\rm osc} &= \frac{1}{2} \sin^22\theta 
  \left[1 - 
  \exp\left(-\frac{\delta^2\Delta^2}{2}\right)
  \cos\left(\Delta L\right)
  \right], \qquad ( \tau \gg \delta) \,.
\end{align}
This formula corresponds  to the usual localization condition: observability  of oscillations requires  that  processes occurring  at the  source and the detector are much  better localized than the neutrino oscillation length, $\delta \ll L_{\rm osc}$.  Consider, for example, a non-relativistic detector particle and assume that it is localized within a distance $\delta_D \sim D$, where  $D$  is a  typical inter-atomic or inter-molecular distance. If the detector particle moves non-relativistically with velocity $\beta_D \ll 1$, it will be localized in time with $\tau_D \sim D/\beta_D \gg \delta_D$. Hence, if for the source we have $\tau_S \gtrsim \delta_S$, we recover the ``stationary detector'' case for non-relativistic detector particles. Another example of $\tau_D \to \infty$ would be a detector particle in a bound state with fixed energy eigenvalue.

There is an interesting consequence of the assumption that the detector is stationary. Indeed,   stationary detector implies  that the  phase-space density $n_D$ does not depend on time.\footnote{As we already mentioned, for the Gaussian model, this is achieved by taking 
$\tau_D \to \infty$.}  
Then the $\d t_D$ integral in \cref{eq:W4} gives a delta-function $\delta(\kappa^0)$ and the $\d \kappa^0$ integral sets $\kappa^0=0$,  which means that interfering neutrinos have identical energies \cite{Grimus:1996av,Grimus:1998uh}. In the approach of \cite{Krueger:2023skk,Akhmedov:2010ms} this is obtained by integrating over an unobservable neutrino propagation time (which resembles somewhat the $\d t_D$ integration in our approach). The assumption of a completely stationary detector has been relaxed in \cite{Grimus:2019hlq}.

\begin{figure}
    \centering
    \includegraphics[width=0.6\linewidth]{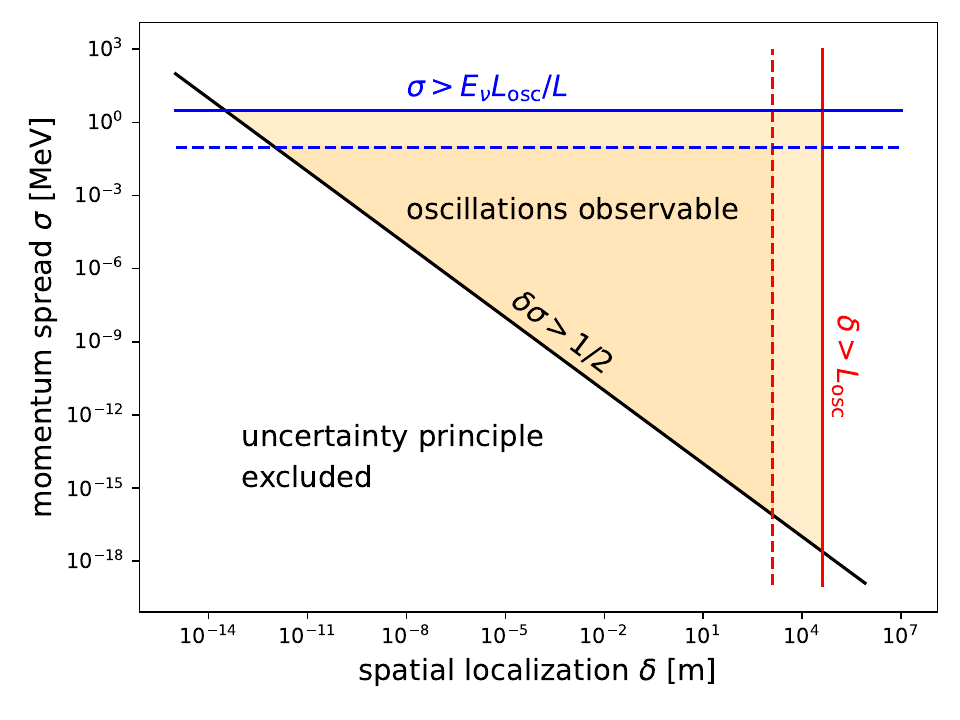}
    \caption{Values of spatial and momentum spreads required for the observability of oscillations (shaded region) from the coherence conditions implied by \cref{eq:damping-all} (red and blue lines) and the uncertainty principle (black line). For illustration purposes we adopt parameters relevant for the JUNO reactor experiment: $E_\nu = 4$~MeV, $L=53$~km. Solid (dashed) lines refer to $\Delta m^2 = 7.5\times 10^{-5} \,(2.5\times 10^{-3})$~eV$^2$.} 
    \label{fig:decoh}
\end{figure}

Moving to the momentum integrals, we can perform the $\d^3\vec l_{S,D}$ integrals using  spatial parts of the delta-functions in \cref{eq:Gamma,eq:sigma}, such that the Gaussian factors become
\begin{align}\label{eq:Gauss2}
  \exp\left[ -\frac{(\vec P_S + \vec p_\nu - \bar {\sf l}_S)^2}{2\sigma_S^2}
    -\frac{(\vec P_D - \vec p_\nu - \bar {\sf l}_D)^2}{2\sigma_D^2}
    \right] \,. 
\end{align}
These factors will have to be 
integrated further over $\d^3\vec p_\nu$. The angular part of this  integral will be constrained by the Gaussian factor
in \cref{eq:Pdecoh-space-time} which involves transverse directions. Then we are left with an integral of the type
\begin{align}
\int \d E_\nu \, h(E_\nu) e^{-g(E_\nu) \pm i\Delta  L} \,,    
\end{align}
where $h(E_\nu)$ is a sufficiently smooth function of $E_\nu$ and
$g(E_\nu)$ is a quadratic function in $E_\nu$ with 
\begin{align}\label{eq:sigma_eff}
  g''(E_\nu) = \frac{1}{\sigma_S^2}+\frac{1}{\sigma_D^2}\equiv\frac{1}{\sigma^2} \,.
\end{align}
Applying  the usual method (see e.g., \cite{Beuthe:2001rc}, or Appendix~A of \cite{Krueger:2023skk}), we find that this integration leads to an additional damping factor in \cref{eq:Pdecoh-space-time} such that the overall decoherence factor in front of the interference term becomes
\be
\begin{split}
    &\exp\left[-\frac{1}{2} \left(\bar\Delta L \frac{ \sigma}{\bar E_\nu}\right)^2 \right]
      \exp\left(-\frac{1}{2} \frac{\tau^2\delta^2}{\tau^2+\delta^2} \bar\Delta^2\right) 
      \quad \to\quad \\
    &\exp\left[-\frac{1}{2} \left(\frac{\Delta m^2 L \sigma}{2\bar E_\nu^2}\right)^2 \right]
    \exp\left[-\frac{1}{2} \left(\frac{\Delta m^2 \delta}{2\bar E_\nu}\right)^2 \right], \qquad ( \tau \gg \delta) \,, \label{eq:damping-all}
\end{split}
\ee
where  $\bar E_\nu$ is the neutrino energy that minimizes  $g(E_\nu)$. 
Hence, within the toy  Gaussian  model, we recover the usual damping factors due to momentum and spatial localizations. Taking into account the fundamental uncertainty relation \cref{eq:QM}, we see that oscillations would disappear, either for $\sigma\to 0$ or $\delta \to 0$, that is, intrinsic delocalisation both in space and momentum are required, as illustrated in \cref{fig:decoh}.

\bigskip

Let us now discuss decoherence effects related to the violation of some assumptions we adopted in \cref{sec:deriv} to  derive the oscillation probability.
A crucial assumption in this derivation was the approximation
that  integrations over the neutrino invariant mass 
$Q^2$ and the 4-vector $\kappa$ in \cref{eq:W5}
can be performed independently of the remaining terms in the integrand. 
While this assumption is natural, deviations from it lead to corrections to the derived formula
and the pattern of neutrino oscillations will be affected.

Terms in the integrand  in 
eq.~(\ref{eq:W5}) that one may consider include 
the  matrix elements squared for the production and detection   processes,  the respective phase spaces, and the distribution functions $n_{S,D}(\vec r, \vec l, t)$
of source and detector particles.
For example, in  our approach, decoherence may occur  if e.g. products 
of these quantities evaluated at  the poles of two different neutrino propagators,  $Q^2 = m^2_1$ and $Q^2 = m_{2}^2$, are so different from each other, that these differences cannot be ignored. 

To provide an example, we will focus on the dependence of the distribution functions $n_X(\vec r_X, \vec l_X, t_X)$ ($X=S,D$) on the momenta $\vec l_X$. To be specific we consider the source particle $X=S$, but the discussion is fully analogous for the detector.
Since the main contribution to the integral over $Q^2$ comes from poles of the neutrino propagators in the  two diagrams in \cref{fig1}, and since we keep momenta of final state particles fixed, the momentum of  the  decaying particle  $S$ will be different in the two cases. Starting from the equation
\be
(l_{S_{1,2}} - P_{S})^2 = m_{1,2}^2,
\ee
and writing 
$l_{S_1} = (E_{S_1}, \vec l_S)$ and $l_{S_2} = (E_{S_2}, \vec l_S + \delta \vec l_S)$, it is easy to derive an estimate of the momentum differences between the residues for the two neutrino propagators
\be
|\delta \vec l_S| \sim \frac{ \Delta m^2}{|\vec Q|} \sim \frac{1}{L_{\rm osc}} \,.
\label{eq45}
\ee
The standard pattern of neutrino oscillations 
is obtained if the difference between distributions computed at $\vec l_S$ and $\vec l_S + \delta \vec l_S$
can be neglected 
\be
| n_S(\vec r_S, \vec l_S + \delta \vec l_S, t_S)
- n_S(\vec r_S, \vec l_S, t_S)| \ll n_S(\vec r_S, \vec l_S, t_S) \, .
\label{eq46}
\ee
Expanding the left-hand side of \cref{eq46} in $\delta \vec l_S$, 
we find 
\be
\frac{\partial n_S(\vec r_S,\vec l_S,t_S)}{\partial \vec l_S} \; \delta \vec l_S  
\sim \left|\frac{\partial n_S(\vec r_S,\vec l_S,t_S)}{\partial \vec l_S} \right| \; L_{\rm osc}^{-1} \ll n_S(\vec r_S, \vec l_S ,t_S) \,.
\ee
We estimate  the derivative of the density using the inverse momentum spread $1/\sigma_S$ to arrive at the condition
\be
\frac{1}{\sigma_S L_{\rm osc}} \ll 1 \,.
\ee
This can be compared to the localisation condition $\delta \ll L_{\rm osc}$, see \cref{eq:decoh-stationary}. Together with the uncertainty relation \cref{eq:QM} we see that
\be
\frac{1}{\sigma_X L_{\rm osc}} < 
\frac{2 \delta_X}{L_{\rm osc}} \ll 1 \qquad (X=S,D) \,,
\ee
where we generalized to source and detector. We see that the localization condition derived above by the classical averaging of the oscillation probability implies that the approximations adopted to perform the integrations in our derivation are automatically satisfied. This is another example of the fact that  classical averaging and intrinsic quantum-mechanical decoherence effects are indistinguishable \cite{Kiers:1995zj,Stodolsky:1998tc,Ohlsson:2000mj,Krueger:2023skk}.

%%%%%%%%%%%%%%%%%%%%%%%%%%%%%%%%%%%%%%%%%%%%%%%%%%%%%%%%%%%%%%%%%%%%%%%%
\section{Effects of the decaying  particle in the neutrino source}
%%%%%%%%%%%%%%%%%%%%%%%%%%%%%%%%%%%%%%%%%%%%%%%%%%%%%%%%%%%%%%%%%%%%%%%%
\label{sec:resonance}

\subsection{Derivation}

We now extend the discussion 
in section~\ref{sec:deriv}
and consider  the case when  the neutrino in the source is produced by a decaying particle $R$, see refs.~\cite{Grimus:1999ra,Akhmedov:2010ms,Grimus:2019hlq}
for earlier discussions of this problem.  This particle  
is produced in  a process $1+2 \to R + f_R$, where $f_R$ is a final state with fixed momentum $P_{f_R}$. The resonance $R$  decays producing  internal $t$-channel neutrino, $R\to \nu + f_S$. 

We note that in addition to the academic interest, the discussion of a resonance is  useful for understanding  a set up where neutrinos are produced in  decays of  relativistic   pions  in a collimated beam,   or for reactor neutrinos, where the resonance is a $\beta$-decaying nucleus produced in the fission of a mother nucleus.

The overall process is now $1+2+3 \to f_R + f_S + f_D$, where we gave  the detector particle $D$ a label $3$ for notational convenience. The $S$-matrix element of the total process is
\begin{align}\label{eq:Sresonance}
  S_{if} = \int \prod_{i=1}^3  
  [\d \vec k_i] \;   \phi_i(\vec k_i)
  (2\pi)^4\delta^{(4)}(k_{123} - P_{f_R}-P_S-P_D) A(k_1,k_2) W(q_R) \mathcal{M}_{if} \,,
\end{align}
where $A$ is the amplitude of  the process $1+2\to R+f_R$, 
\begin{equation}
 W(q_R) = \frac{1}{q_R^2 - M^2 + iM\Gamma}
\end{equation}
is the Breit-Wigner propagator,\footnote{The quantum numbers of the resonance are not relevant; for this reason we assumed it to be a scalar particle.}  $q_R = k_{12} - P_{f_R}$ and $\mathcal{M}_{if}$ is the amplitude for the process $R+D \to f_S+f_D$. We use the short-hand notations $k_{12}$, $k_{123}$ to denote  sums over the corresponding momenta. Furthermore, we assume that momenta $k_{1,2}$ are such that the on-shell production of the  resonance $R$  is possible, i.e. that the condition  
$(k_{12}-P_{fR})^2 = M^2$
is kinematically accessible.

To compute the probability
of the process 
$1+2+3 \to f_R + f_S + f_D$, we square the amplitude $S_{fi}$. Upon doing this,  similar to the discussion in  Section~\ref{sec:deriv},  we obtain two delta-functions, one with the argument 
$k_{123} - P_{f_R}-P_S-P_D$ and another one with the argument 
$k'_{123} - P_{f_R}-P_S-P_D$. 
Following the discussion in Section~\ref{sec:deriv}, we rewrite these $\delta$-functions 
 by introducing momenta for the resonance $R$ and the $t$-channel neutrino  
\be
\begin{split}
  &\delta^{(4)}(k_{123} - P_{fR}-P_S-P_D) = \\
  &\int \d^4q_R \d^4q_\nu \,
  \delta^{(4)}(k_{12} - q_R - P_{fR})
  \delta^{(4)}(q_R - q_\nu - P_{S})
  \delta^{(4)}(k_3 + q_\nu - P_{D}) \,.
\end{split}
\ee
We do the same with the ``primed'' momenta and then move on to two average four-momenta and their differences, similar to  what is done in \cref{eq:momenta}
\begin{equation}
\begin{split}
& Q_R = \frac{q_R + q_R'}{2}, 
\;\;\; \kappa_R = q_R - q_R' \,,
\\
& Q\;\; = \frac{q_\nu + q_\nu'}{2},
\;\;\;\;\; \kappa \;\; = q_\nu - q_\nu'\,.
\end{split}
\end{equation}
The product of two delta-functions that appears in $\d W_{if}$ becomes
\be
\begin{split}
\int \d^4Q_R\d^4Q \d^4\kappa_R\d^4\kappa\,  
&\delta^{(4)}(l_{12} - Q_R - P_{fR})
  \delta^{(4)}(Q_R - Q - P_{S})
  \delta^{(4)}(l_3 + Q - P_{D})  \\
  &\times \delta^{(4)}(\kappa_{12} - \kappa_R)
  \delta^{(4)}(\kappa_R - \kappa)
  \delta^{(4)}(\kappa_3 + \kappa_\nu) \,.  \label{eq:deltaQ_R}
\end{split}
\ee
The transition probability $\d W_{if}$ can  be calculated 
following the discussion in Section~\ref{sec:deriv}. We provide the details of this calculation in \cref{app:resonance}. If follows from that discussion that the integral $I$
in eq.~(\ref{eq2.26}) -- a quantity that is directly related to  the neutrino oscillation probability -- is generalized to 
\begin{equation}
\begin{split}
I_R = 2 \Gamma M\;
\int  &\frac{{\rm d} Q_R^2}{2 \pi}
\frac{{\rm d} Q^2}{2 \pi}
\frac{{\rm d}^4 \kappa}{(2\pi)^4}
e^{-i \kappa_\mu x^\mu} \,
 W\left (Q_R + \frac{\kappa}{2} \right ) W^*\left (Q_R - \frac{\kappa}{2} \right  ) \\
%F_\nu(Q+\frac{\kappa}{2} ) 
%F^*_{\nu}(Q-\frac{\kappa}{2}),
&\times \sum_{a,b=1,2} \frac{(-1)^a(-1)^b}{(Q^2 + Q\kappa -m_a^2+i\epsilon)(Q^2-Q\kappa-m_b^2-i\epsilon)} \,,
\label{eq:IR}
\end{split}
\end{equation}
where $x^\mu = (t_D - t_S, \vec r_D - \vec r_S) $. 
We note that $t_S$ and $\vec r_S$ in this case refer to the 
space-time point where particles $1$ and $2$ collide to produce a resonance $R$, rather than the space-time point where the neutrino was produced.

Equation~(\ref{eq:IR}) describes two effects -- the impact of the finite lifetime of a resonance on the neutrino oscillation probability, and the effect of the smearing of the neutrino energy caused by the resonance  off-shellness. Both of these effects can be accounted for at once but, for the sake of simplicity we will focus on the former and then briefly comment on the latter. 

To describe  the impact of the resonance finite lifetime, we 
assume  that the neutrino  energy is independent of the resonance off-shellness.
This allows us to integrate over $Q_R^2$ in eq.~(\ref{eq:IR}).
The Breit-Wigner terms have poles at
\begin{equation}\label{eq:polesQR}
    Q_R^2 = M^2 \pm \kappa  Q_R \pm i M\Gamma \,,
\end{equation}
and, applying the residue theorem, we find 
\begin{equation}
\int \frac{{\rm d} Q_R^2}{2 \pi} \,
W\left(Q_R + \frac{\kappa}{2} \right) 
W^*\left(Q_R - \frac{\kappa}{2} \right)
= \frac{1}{2 Q_{R}^0} \;  \frac{i}{
\kappa \beta_R + i \Gamma/\gamma_R} \,,
\end{equation}
where $\beta_R^\mu = (1, \vec \beta_R)$ is the four-velocity of the resonance and $\gamma_R = Q_{R}^0/M$ is the resonance $\gamma$-factor. 
To perform subsequent integrations over $Q^2$ and $\kappa$, we write
\begin{equation}
\frac{i}{
\kappa \beta_R + i \Gamma/\gamma_R}
= \int \limits_{0}^{\infty}
{\rm d} t_R \; e^{-t_R/\tau_R} \; 
e^{i \kappa \beta_R t_R} \,,
\end{equation}
where $\tau_R = \gamma_R/\Gamma$ 
is the lifetime of the resonance in the laboratory frame.  Using this representation in \cref{eq:IR}, we observe that the integration over $Q^2$ and $\kappa$ can be performed following what was discussed earlier in 
\cref{sec:deriv}. We obtain 
\begin{equation}
I_R = 
\frac{1}{ \tau_R} 
\int \limits_{0}^{\infty} 
{\rm d} t_R \;
e^{-t_R/\tau_R} \; 
I(x_D^\mu - x_S^\mu - \beta_R^\mu t_R,Q) \,,
\label{eq4.10}
\end{equation}
where $I$ is given in \cref{eq:I}.
\Cref{eq4.10} has a clear physical interpretation, namely the
smearing of the distance, over which oscillations are observed, over the 
finite life-time of the resonance. 
To quantify this  effect, we consider the following integral
\begin{align}
    I_R &= 
\frac{2}{\tau_R  |\vec Q|}
\int \limits_{0}^{\infty} 
{\rm d} t_R \, e^{-t_R/\tau_R}
\sin^2 \frac{( r_{||} - \beta_{R||} t_R)}{L_{\rm osc}}
 \,,
 \label{eq4.11}
\end{align}
which follows from \cref{eq4.10} 
if the transverse $\delta$-functions and their dependence on $\vec \beta_R$ are neglected. As
in \cref{sec:deriv}, we use the notation  $r_{||} = (\vec r_D  - \vec r_S) \cdot \vec e_\nu$, 
and $\beta_{R ||} = \vec \beta_R \cdot \vec e_\nu$.

The integration in 
eq.~(\ref{eq4.11}) is straightforward, and we find 
\be
I_R = \frac{1}{
 |\vec Q|
} 
\left [ 1 - \cos \xi
\cos \left (  \frac{2 r_{||}}{L_{\rm osc}} - \xi 
 \right )
\right ] \,,
\label{eq4.12}
\ee
where 
\be
\tan \xi \equiv 2  \, \frac{\beta_{R ||}\tau_R}{L_{\rm osc}} = \frac{2Q_{R||}}{L_{\rm osc}M\Gamma} \,.
\label{eq4.13}
\ee
It follows that a modification of the standard neutrino oscillation probability formula occurs if the path travelled by the resonance before it decays is comparable to the oscillation length
\be
\beta_{R ||} \tau_R 
\sim L_{\rm osc}.
\ee
\Cref{eq4.12} agrees with eq.~(31) of ref.~\cite{Akhmedov:2012uu} in the limit of an infinitely long pion-decay tunnel ($l_p \to \infty$). In ref.~\cite{Akhmedov:2012uu} this expression has been obtained from a quantum-mechanical description of neutrino oscillations based on  wave-packets.

We now return to \cref{eq:IR} and discuss the second effect of the resonance finite lifetime  on neutrino oscillations, namely the dependence of the neutrino energy $|\vec Q|$ on the resonance  off-shellness $Q_R^2$.
In this case, the computational strategy described above breaks down as we cannot easily integrate over $Q_R^2$. However, it is possible to integrate over $Q^2$ 
and $\kappa$ first, 
and over $Q_R^2$ later.   We do not discuss in detail   how to do this and, instead, explain how to obtain a good approximation for the final result using  plausible  physical considerations. 

To this end, we note that the  dependence of the neutrino energy on the invariant mass of the resonance can be written as follows 
\be
E_\nu(Q_R^2) = {\bar E}_\nu
+ \frac{\partial E_\nu}{\partial Q_R^2} (Q_R^2 - M^2) 
+ \cdots 
\ee
where ellipses stand for higher order terms in the expansion in $Q_R^2 - M^2$.
Upon integrating over 
$Q_R^2$, $Q_R^2 - M^2$ receives an  imaginary part $ \pm i M \Gamma$, see \cref{eq:polesQR}. Since the oscillation length
in \cref{eq4.12} is  proportional to
 $E_\nu$, 
the shift in the neutrino energy leads to the following change in the neutrino oscillation probability
\be
P_{\rm osc} 
= \frac{1}{2} 
\sin^2 2 \theta
\left [ 
1 - e^{-\frac{2  r_{||}}{L_{\rm osc}}
\frac{\mu \Gamma}{M}
}
\cos \xi \cos 
\left (\frac{2 r_{||}}{L_{\rm osc}} - \xi  \right )
\right ],
\label{eq4.15}
\ee
where 
\be
\mu = 
\frac{M^2 }{E_\nu} \frac{\partial E_\nu}{\partial Q_R^2} \sim 1,  
\ee
is a parameter that characterizes the decay at the source.\footnote{Let us note that the derivative $\partial E_\nu / \partial Q_R^2$ is determined by 
the kinematical constraint $\delta^{(4)}(Q_R - Q - P_S)$ from \cref{eq:deltaQ_R}: 
neglecting $Q^2 \approx m_\nu^2$, it implies $Q_R^2 \approx 2Q P_S + P_S^2 \approx 2E_\nu \tilde P_S  + P_S^2$ with
$\tilde P_S \equiv P_S^0 - \vec P_S \cdot \vec e_\nu$, and therefore 
$\partial E_\nu / \partial Q_R^2 = 1/(2\tilde P_S)$.}

%%%%%%%%%%%%%%%%%%%%%%%%%%%%%%%%%%%%%%%%%%%%%%%%%%%%%%%%%%%%%%%%%%%%%%%%
\subsection{Discussion} 

\Cref{eq4.15} shows that neutrino 
oscillations are damped for both, too small and too large decay width of the resonance. If  
$\Gamma \ll 1/L_{\rm osc}$, then $\xi \to \pi/2$,  $\cos \xi$ in 
eq.~(\ref{eq4.15}) vanishes and oscillations disappear because of the integration over the path travelled by the resonance. If, on the other hand, the width is such that 
$\Gamma \gg  M L_{\rm osc}/r_{||}$, the oscillatory pattern will be exponentially suppressed. In this case the reason is the large energy spread induced by the off-shellness of the resonance. %Combining the two %conditions implies 
For the observability of oscillations, these two conditions imply the following constraint on the resonance width 
%
%\be
%\frac{Q_{R||}}{M\Gamma L_{\rm osc}} \ll \Gamma \ll M \frac{L_{\rm osc}}{r_\parallel} \,.
%\ee
\be
\frac{\Delta m^2}{M} \ll \Gamma \ll M \,,
\label{eq4.18}
\ee
where we have assumed $Q_{R\parallel} \sim E_\nu$ and 
that neutrinos are  detected around the oscillation maximum, $r_{||} \sim L_{\rm osc}$. For typical parameter values, there are many orders of magnitude available for $\Gamma$ to fulfil this condition.

As noted in \cite{Akhmedov:2012uu}, the condition $\beta_{R||}\tau_R \ll L_{\rm osc}$ implies an upper bound on the mass-squared difference 
for which oscillations can still be observed.
To illustrate this point,  consider an example of a neutrino beam produced in  pion decays. From eq.~(\ref{eq4.13}) and the requirement $\xi \ll 1$, we find
\be
\Delta m^2  \ll  \frac{2E_\nu M_\pi \Gamma_\pi}{p_\pi} \,,
\ee
where $M_\pi,\Gamma_\pi,p_\pi$ are the pion mass, its decay width, and its  momentum, respectively. For numerical estimates, we consider a neutrino beam with an off-axis angle $\alpha$ with respect to the pion direction, with $\alpha \ll 1$ and $p_\pi \gg M_\pi$. Then we obtain from kinematics the approximate expression  
\begin{align}
\Delta m^2  \ll  \frac{M_\pi^2\Gamma_\pi}{E_{\nu 0}} 
\left(1 - \frac{E_\nu^2}{4E_{\nu 0}^2}\alpha^2\right)  
\approx 3\,{\rm eV^2}
\left(1 - \frac{E_\nu^2}{4E_{\nu 0}^2}\alpha^2\right) \,,
\end{align}
where $E_{\nu 0} = (M_\pi^2-M_\mu^2)/(2M_\pi) \approx 30$~MeV is the neutrino energy in the pion rest frame. This constraint is relevant for sterile neutrino searches in neutrino beams from pion decays, and shows that for mass-squared differences of order eV$^2$ the effects of the resonance path-length will start affecting the observability of the oscillatory pattern. 
\\

In this section we have derived the intrinsic smearing effects due to the finite lifetime of the particle producing the neutrino. In addition to these effects, the localization in space, momentum and time of the effective phase-space density for the resonance defined in \cref{eq:n_R} 
impacts the oscillations in 
a similar way as discussed in \cref{sec:decoh}. At leading approximation, the smearing effects factorise and the damping factors  derived in \cref{sec:decoh} will appear in addition to the effects discussed above. If the resonance momentum $\vec Q_R$ is smeared with a Gaussian with the width $\sigma_R$, two effects arise. First, the smearing of $\vec Q_R$ affects the neutrino momentum $\vec Q$ due to the delta-function $\delta^{(3)}(\vec Q_R - \vec Q -\vec P_S)$ with $\vec P_S$ fixed, and second, there is a direct effect of smearing $\vec Q_R$ via the term $\beta_\parallel\tau_R = Q_{R\parallel} /(\Gamma M)$ in the effective neutrino distance. At leading order, they lead to a damping factor 
\begin{align}
\exp\left[-\frac{1}{2}\left(\Delta L \frac{\sigma_R}{E_\nu}\right)^2\left(1 + \frac{E_\nu}{LM\Gamma} \right)^2  
\right] \,.  
\end{align}
The term $(\Delta L \sigma_R/E_\nu)^2$ is the same as in \cref{eq:damping-all} and originates from the smearing of the neutrino energy, whereas the term $E_\nu/(LM\Gamma)$ comes from the smearing of the effective distance between the source and the detector. Unless $\Gamma$ is very small, this term typically is sub-leading.

As a further remark, we note that in  this section we have assumed that the resonance travels in vacuum and does not interact with particles in the environment. If such interactions are important, there will be energy and momentum exchanges of the resonance with the environment before decay, and the treatment with the free vacuum propagator used here would not apply. 
If the interaction rate with the environment is much faster than the decay rate, we can treat the resonance as an effective on-shell particle, whose phase-space density is determined by its interaction with environment and is given by $n_S(\vec r_S, \vec l_S, t_S)$ as discussed in \cref{sec:deriv}. Hence, in such a case the results of \cref{sec:deriv} apply. 

Regarding specific experimental realisations, the  calculation described in  this section  applies to the case of a 
pion-based neutrino  beam, where pions decay in flight within a decay pipe, such as for instance in the T2K, NOvA, MiniBooNE or future DUNE and T2HK experiments. We have neglected, however, the case when the decay pipe is shorter than (or comparable to) a  typical path travelled by the pion, see ref.~\cite{Akhmedov:2012uu} for a discussion. 
It is 
straightforward to include the  finite length of the decay pipe $l_p$ in our calculations, by introducing a theta-function 
$\theta(l_p - \beta_{R||}t_R)$ to cut-off the integral in \cref{eq4.11}, or generalize it   if the collinear approximation does not hold and  pions may hit  side walls of the pipe before decaying to a neutrino.

In contrast, for reactor experiments  the beta-decaying fission products typically have long lifetimes and thermalise in the reactor core before decaying. In that case, interactions dominate over 
finite lifetime effects \cite{Krueger:2023skk}, and the formalism of \cref{sec:deriv} applies. Similar arguments hold for experiments using pion decay at rest at a stopped pion source, such as the LSND experiment.

%%%%%%%%%%%%%%%%%%%%%%%%%%%%%%%%%%%%%%%%%%%%%%%%%%%%%%%%%%%%%
\section{Summary and discussion}
%%%%%%%%%%%%%%%%%%%%%%%%%%%%%%%%%%%%%%%%%%%%%%%%%%%%%%%%%%%%%
\label{sec:summary}

We have presented an alternative  method to derive the standard vacuum neutrino oscillation probability from   quantum field theory. Our derivation is based on the methods  applied previously to describe collider processes that involve elementary particles,  yet    are sensitive to geometric sizes of colliding beams \cite{Ginzburg:1991mv,Ginzburg:1991mw,Kotkin:1992bj}. 
Although the  results derived in this paper are generally known, 
the approach to their derivation appears to be very  transparent, providing  further insights into  conditions required for the observability of oscillations. 
A straightforward generalization of the calculation allows us to describe the impact of the decaying particle's finite lifetime on neutrino oscillations.

Our approach enables the  derivation of  the standard oscillation formula without the need to specify the shape of the initial-state wave packets. Instead of working with transition amplitudes, we consider the transition probability (proportional to the amplitude-squared) from the very beginning. These aspects of our derivation are  similar to the one presented in appendix~A of \cite{Falkowski:2019kfn}, arXiv version~v3. There, the interference term is obtained from a pole-integration, which has similarities to our $\d^4 \kappa$ integration, see \cref{eq.2.27}. The latter generalizes the corresponding expression in \cite{Falkowski:2019kfn} to four space-time dimensions, allowing for a consistent discussion of  time dependencies,  as we discuss below.

Following refs.~\cite{Ginzburg:1991mv,Ginzburg:1991mw,Kotkin:1992bj},
we consider a statistical average of the probability which leads to density matrices for neutrino source and detector particles, which in turn are expressed in terms of semi-classical phase-space densities
with the help of the Wigner transformation \cite{Case2008}. This approach offers a very transparent way to take into account properties of neutrino source and detector. One of our main results is \cref{eq:W7}, where the transition rate is obtained as a convolution of the production rate, the oscillation probability and the detection cross section with the source and detector phase-space densities. 

 The phase-space densities offer a direct way to describe the time structure of the  source and the detector. There is no need to perform a somewhat  ad-hoc averaging over detection or production times, as happens in some other approaches,  since time integrals
 arise naturally in our  calculation. 
As the result, this  method naturally covers situations such as  pulsed neutrino beams. The detector phase-space density is evaluated at the ``retarded time'', 
reflecting the propagation time of neutrinos between the source and the detector. The time dependence of the detector is important  for the resulting
oscillation pattern. In the realistic case, where the detector particle is localised in space much better than  in time (in natural units), oscillations will occur in space and not in time. The limiting case corresponds to the ``stationary detector'', where the detector particle phase-space density can be considered constant in time. 

Our approach clearly displays the different roles of spatial and momentum localizations. The spreads $\delta$ in space and $\sigma$ in momentum are independent quantities in general, only subject to the lower bound implied by the uncertainty relation $\sigma \delta > 1/2$. For illustration purposes, in \cref{sec:decoh} we employed a toy   Gaussian  model for the phase-space densities, to derive the well-known coherence conditions $\delta \ll L_{\rm osc}$ and $\sigma \ll E_\nu$. 

The effective uncertainties derived in the toy Gaussian model of \cref{sec:decoh} differ somewhat from 
%the corresponding 
expressions obtained in e.g. \cite{Giunti:1997sk,Beuthe:2001rc,Akhmedov:2010ms} and recently summarized in \cite{Krueger:2023skk}. In the traditional method, a velocity has to be introduced by expanding the dispersion relation and keeping only first-order terms (neglecting wave packets dispersions). Effective spreads depend on the velocity-weighted individual spreads and/or on an effective velocity-squared. In our method we neither need to specify the dispersion relation of initial state particles nor consider its expansion. The relevant space, time and momentum uncertainties are simply either squared-sums or inverse squared-sums of the individual spreads, see \cref{eq:spreads-space-time,eq:sigma_eff}.

A crucial assumption of our method is that all 
final-state particles are exact momentum eigenstates, see also \cite{Grimus:1996av, Falkowski:2019kfn, Grimus:2019hlq}. The model of ref.~\cite{Beuthe:2001rc} used e.g., in ref.~\cite{Krueger:2023skk} would lead to a vanishing interference term, in the case of a single source and detector particle and sharp final-state momenta. We believe that this is a consequence of expanding the dispersion relation only up to linear order combined with the Gaussian wave-packet ansatz, which leads to an intricate cancellation of terms setting the effective energy spread to zero. This is different from our model, which allows for interference under these assumptions. We have checked indeed that kinematics does not fully determine the neutrino momentum under these conditions, and therefore, an interference term is not in contradiction with the energy-momentum conservation implied by the sharp final-state momenta. 
   
The last item may have also phenomenological consequences. Contrary to our assumption that  momenta of final-state particles are fixed,  in the approach of refs.~\cite{Giunti:1993se, Beuthe:2001rc, Akhmedov:2010ms, Krueger:2023skk} their uncertainties are taken into account. 
It is found that these uncertainties impact the coherence properties. In fact, it  turns out, that  the effective energy-smearing is often dominated by the outgoing charged lepton, which typically has a large mean-free path and therefore a fairly well-defined  momentum. This  leads to values of $\sigma$ which can be ${\cal O}(10^3)$ times smaller than what one obtains by considering  initial-state particles, see e.g.  table~1 in  ref.~\cite{Krueger:2023skk}, or ref.~\cite{Akhmedov:2022bjs} for similar results based on the wave-packet arguments. Our assumption of fixed 
final-state momenta is rooted in the standard particle physics calculation of scattering rates and cross sections. We leave the question of whether for certain experimental configurations this approach needs to be adapted to correctly describe phenomenology of neutrino oscillations for further studies.  

In contrast, the spatial localization $\delta$ from \cref{eq:spreads-space-time} is of a similar size as the corresponding quantity $1/\sigma_m$ in \cite{Krueger:2023skk}, which also turns out  to be dominated by properties of the 
initial-state particles in that formalism.
We note that additional smearing effects both in the initial and final states  can be straightforwardly included  by (classically) integrating the fully differential quantity $\d W_{if}$ over (some of) the final-state momenta, as well as over initial-state 
phase-space densities.

Finally, as a straightforward generalization
of our method, we have discussed the implication of the finite width of the decaying particle producing the neutrino in the source. We have identified two effects that may affect the oscillation pattern.  First, in the case of a ``small'' 
decay width,  the uncertainty in the exact displacement of the  resonance before it decays may lead to damping of oscillations, if the resonance mean path becomes comparable to the oscillation length. Second, for a  ``large'' decay width, the off-shellness of the resonance  implies the  energy spread 
of produced neutrinos which can lead to an exponential damping of the interference term. Typically this effect is negligible, as long as $\Gamma \ll M$. Nevertheless, it is interesting that neutrino oscillations disappear in both limits, $\Gamma \to 0$ and  $\Gamma \to \infty$, and 
the condition for their observability is given in eq.~(\ref{eq4.18}). This equation is typically satisfied for practical cases, so that oscillations are unaffected by resonance effects, except for sterile neutrino searches with $\Delta m^2 \gtrsim 1$~eV$^2$ at pion beam experiments. 
Finally, we note that our treatment of the resonance in \cref{sec:resonance} applies to cases, when interactions of the decaying particle with the environment can be neglected. This is the case for the experimentally important configuration of a neutrino beam from the pion decay in flight.

{\acknowledgments{K.M.\ is  grateful to  
D.V.~Naumov, V.~Serbo and A.~Vainshtein for useful conversations and comments on the manuscript. K.M.\ would like to dedicate this paper to the memory of I.F.~Ginzburg and G.L.~Kotkin, from whom he learned about the calculational  method used in   this paper,  and about  life in  theoretical physics in a more general sense. 
This research was supported in part by grant NSF PHY-2309135 to the Kavli Institute for Theoretical Physics (KITP), by  
Deutsche Forschungsgemeinschaft (DFG, German Research Foundation) under grant no.\ 396021762 - TRR 257, and by the European Union’s Framework Programme for Research and Innovation Horizon 2020 under grant H2020-MSCA-ITN-2019/860881-HIDDeN.
}}
%%%%%%%%%%%%%%%%%%%%%%%%%%%%%%%%%%%%%%%%%%%%%%%%5
\appendix

\section{Resonance calculation}
\label{app:resonance}

In this appendix, we 
provide additional details about  the calculation of  the transition probability $\d W_{if}$ for the case of a resonance  production and decay in the neutrino source,  discussed in \cref{sec:resonance}. The required manipulations are analogous  to the ones described in \cref{sec:deriv}.
Starting with  the  $S$-matrix element  in \cref{eq:Sresonance} and using the representation of the  product of delta-functions from \cref{eq:deltaQ_R},  we obtain the following expression for the transition probability
\begin{equation}
\begin{split} 
 \d W_{if} = 
\int & \prod \limits_{i=1}^{3} 
[{\rm d} \vec k_i] [{\rm d} \vec k_i'] \phi_i(\vec k_i) \phi_i^*(\vec k_i') \; {\rm d} \Phi_{f_R} {\rm d} \Phi_S 
{\rm d} \Phi_D 
\frac{{\rm d}^4 Q_R}{(2\pi)^4}
\frac{{\rm d}^4 Q}{(2\pi)^4}
\frac{{\rm d}^4 \kappa_R}{(2\pi)^4}
\frac{{\rm d}^4 \kappa}{(2\pi)^4}
\\
& \times (2 \pi)^4 \delta^{(4)}(Q_R + P_{fR}
- l_{12} ) (2\pi)^4 \delta^{(4)}(k_{12}
- k_{12}'- \kappa_R)
\\
& \times (2 \pi)^4 \delta^{(4)}(Q + P_S - Q_R) (2\pi)^4 \delta^{(4)}(\kappa_R - \kappa)
\\
& \times (2 \pi)^4 \delta^{(4)}(P_D - 
Q-l_3) (2\pi)^4 \delta^{(4)}(k_{3}
- k_{3}'+ \kappa)
\\
& \times A(k_1,k_2) A^*(k_1',k_2')
W(Q_R + \kappa_R/2)W^*(Q_R-\kappa_R/2) \; \mathcal{M}_{fi} \mathcal{M}_{fi'}^*,
\end{split}
\end{equation}
where $l_i =(k_i+k_i')/2$ and $\kappa_i = k_i - k'_i$.
%
%To simplify the above equation, we change variables 
%\begin{equation}
%\vec l_i = \frac{\vec k_i + \vec k_i'}{2},
%\;\;\; \vec \kappa_i = \vec k_i - \vec k_i', \;\;\; i = 1,2,3,
%\end{equation}
%and note that with linear ${\cal O}(\kappa_i)$ accuracy, $l_i$ can be considered a four-vector in a sense that 
%\begin{equation}
%l_i^{(0)} = \sqrt{\vec l_i^2 + m_i^2} + {\cal O}(\kappa_i^2).
%\end{equation}
Writing the time-component of the delta-functions involving $\kappa$ as
\begin{equation}
\begin{split}
& 2 \pi \;\delta(k_{12}^0 - k'^0_{12}-\kappa^0) 
= \int \limits_{-\infty}^{\infty}
{\rm d} t_S \; e^{i (E_1' + E_2' - E_1 - E_2+\kappa^0) t_S},
\\
& 2 \pi \;  \delta(k_{3}^0 - k'^0_{3}+\kappa^0) \; 
= \int \limits_{-\infty}^{\infty}
{\rm d} t_D \; e^{i (E_3' -E_3 -\kappa^0) t_D},
\end{split}
\end{equation}
we perform the statistical averaging 
as described in the \cref{sec:deriv}, and write the ensuing density matrices as integrals of the Wigner functions.  Neglecting 
the dependences of amplitudes $A$ on $\kappa_i$ and factorising the  amplitudes $\mathcal{M}_{fi}$ and $\mathcal{M}_{fi' }$ into a product of the corresponding neutrino propagator and ``production'' and ``detection''  
amplitudes that do not depend on $\kappa$, we find 
\begin{equation}
\begin{split}
{\rm d} W_{if} 
& = \int \prod_{i=1}^{3} 
\frac{{\rm d}^3 \vec l_i}{(2\pi)^2 2E_i} {\rm d}^3 \vec r_i {\rm d}^3 \vec \kappa_i e^{-i \vec \kappa_i \cdot \vec r_i}
\; {\rm d} t_S \; {\rm d} t_D
e^{i \kappa_0 (t_S - t_D)}
\; {\rm d} \Phi_S \; {\rm d} \Phi_D 
{\rm d} \Phi_{fR} 
\\
& \times 
\frac{{\rm d}^4 Q_R}{(2\pi)^4}
\frac{{\rm d}^4 Q}{(2\pi)^4}
\frac{{\rm d}^4 \kappa}{(2\pi)^4}
\; n_1(\vec r_1, \vec l_1, t_S)
n_2(\vec r_2, \vec l_2, t_S)
n_3(\vec r_3, \vec l_3, t_D) \; |A(l_1,l_2)|^2
\\
& \times 
(2\pi)^3 \delta^{(3)}( \vec \kappa_{12} - \vec \kappa) 
(2\pi)^3 \delta^{(3)}(\vec \kappa_3 + \vec \kappa )
\;  (2\pi)^4 \delta^{(4)}(Q_R + P_{fR} - l_{12}) 
\\
& \times (2\pi)^4( Q+P_S - Q_R) 
(2\pi)^4(P_D - Q-l_3) |M_{S}|^2 |M_{D}|^2 
\\
&  \times 
 \sin^2 \theta
\cos \theta^2 \; W(Q_R + \kappa/2) W^*(Q_R - \kappa/2) 
F_\nu(Q+\kappa/2) 
F^*_{\nu}(Q/2-\kappa ) \,,
\label{eqa.3}
\end{split}
\end{equation}
where
\begin{equation}
 F_\nu(q) \equiv \sum_{a=1,2} \frac{(-1)^a}
 {q^2 - m_a^2 + i\epsilon} \,.   
\end{equation}
To further simplify eq.~\ref{eqa.3}, we integrate over $\vec \kappa_1$ and $\vec \kappa_2$, 
making use of the fact that the non-trivial part of the integrand only depends on $\vec \kappa_1 + \vec \kappa_2$. This gives a delta-function, which  sets  $\vec r_1 = \vec r_2 \equiv \vec r_S$. Furthermore, we write 
\begin{equation} 
\frac{{\rm d}^4 Q}{(2 \pi)^4 }
= \frac{{\rm d} Q^2}{(2 \pi) }
\frac{{\rm d}^3 Q}{(2\pi)^3 2 Q^0}
\,,\qquad
\frac{{\rm d}^4 Q_R}{(2 \pi)^4 }
= \frac{{\rm d} Q_R^2}{(2 \pi) }
\frac{{\rm d}^3 Q_R}{(2\pi)^3 2 Q^0_R},
\end{equation}
and obtain 
\begin{equation}
\begin{split}
{\rm d} W_{if} 
= \int & \prod \limits_{i=1}^{3} 
\frac{{\rm d}^3 \vec l_i}{2E_i} {\rm d}^3 \vec r_S {\rm d}^3 \vec r_D  \, {\rm d} t_S {\rm d} t_D
\; (2 M \Gamma)^{-1}  |A(l_1,l_2)|^2 {\rm d} \Phi_{fR} 
\frac{{\rm d}^3 Q_R}{(2\pi)^3 2 Q^0_R} 
\\
& \times (2 \pi)^4 \delta^{(4)}(Q_R + P_{fR}
- l_{12}) \, n_1(\vec r_S, \vec l_1, t_S) n_2 (\vec r_S, \vec l_2, t_S)
\\ 
& 
\times |M_{S}|^2 (2\pi)^4 \delta^{(4)}
(Q_R - Q - P_S)
{\rm d} \Phi_S \, \frac{{\rm d}^3 Q}{(2 \pi)^3 2 Q^0}
\\
& 
\times |{\cal M}_{D}|^2
(2 \pi)^4 \delta^{(4)}(Q+l_3 - P_D) 
{\rm d} \Phi_D \, 
n_3(\vec r_D, \vec l_3, t_D) 
\times I_R(Q_R, Q, x),
\end{split} 
\end{equation}
where $I_R$ is given in \cref{eq:IR}. Note that the densities $n_1$ and $n_2$ are evaluated at the same location $\vec r_S$ and at the same time $t_S$ (particles 1 and 2 have to meet at a space-time point to produce a resonance). This suggests the following  definition of an effective phase-space density of the resonance 
\begin{align}\label{eq:n_R}
  n_R(Q_R, \vec r_S, t_S) \propto \int \d^3l_1\d^3l_2  \,
  n_1(\vec l_1, \vec r_S, t_S)
  n_2(\vec l_2, \vec r_S, t_S) \, |A|^2 \,
  \delta^{(4)}(l_{12} - Q_R - P_{fR}) \,.
\end{align}

\bibliographystyle{JHEP_improved}
\bibliography{./refs}

\end{document}